\documentclass[aps,prd,groupedaddress,showpacs]{revtex4}
\usepackage{graphicx}% Include figure files
\usepackage{dcolumn}% Align table columns on decimal point
\usepackage{bm}% bold math
\begin{document}

\title{Shape of the \lowercase{\boldmath{$f_0(980)$}} in
\boldmath{$\gamma\gamma\to\pi^+\pi^-$}}

\author{N.~N.~Achasov}
\email[]{achasov@math.nsc.ru}
\author{G.~N.~Shestakov}
\email[]{shestako@math.nsc.ru}
%\thanks{}
\altaffiliation{} \affiliation{Laboratory of  Theoretical Physics,
S.~L.~Sobolev Institute for Mathematics, 630090, Novosibirsk,
Russia}

\date{\today}

\begin{abstract}

The Belle Collaboration results on the observation of the
$f_0(980)$ resonance in the reaction $\gamma\gamma\to\pi^+\pi^-$
are analyzed. It is argued that they point to the presence of
mechanisms which give rise to a strong distortion of the
$f_0(980)$ resonance shape in comparison with the shape of a
solitary Breit-Wigner resonance. It is shown that the main factors
responsible for the formation of the specific, steplike, shape of
the $f_0(980)$ resonance in the $\gamma\gamma\to\pi^+\pi^-$
reaction cross section are the $K^+K^-$ loop mechanism of the
$f_0(980)$ coupling to the $\gamma\gamma$ system and the
destructive interference between the background and $f_0(980)$
resonance contributions in the $\pi^+\pi^-$ invariant mass region
below the $K^+K^-$ threshold.
\end{abstract}

\maketitle

\section{Introduction}

Recently, the Belle Collaboration succeeded in observing a clear
manifestation of the $f_0(980)$ resonance in the reaction
$\gamma\gamma\to\pi^+\pi^-$ \cite{MBC}. This has been made
possible owing to the huge statistics and good energy resolution.
Evidence for the $f_0(980)$ production in $\gamma\gamma$
collisions obtained previously by the Mark II \cite{E1}, CELLO
\cite{E2}, ALEPH \cite{E3}, Crystal Ball \cite{E4,E5}, and JADE
\cite{E6} Collaborations was essentially less conclusive
\cite{MBC,BP}. The Belle data \cite{MBC} corresponding to the
$f_0(980)$ resonance region are shown in Fig. 1. Figure 1(a) shows
the distribution of $e^+e^-\to e^+e^-\pi^+\pi^-$ and $e^+e^-\to
e^+e^-\mu^+\mu^-$ events, $\Delta N$, in the invariant mass of the
$\pi^+\pi^-$ and $\mu^+\mu^-$ systems, $m$, scanned with a
5-MeV-wide step. A distinct peak due to the $f_0(980)$ resonance
production in the $\gamma\gamma\to\pi^+\pi^-$ channel can be seen
in this plot. The peak position $m_{f_0}=981.2\pm0.5$ MeV and its
total width $\Gamma=21.7\pm2.1$ MeV were determined in Ref.
\cite{MBC} by fitting the $m$ dependence of $\Delta N$ in the
$f_0(980)$ region to the incoherent sum of the resonance and
background contributions:
\begin{equation}
\Delta N=\frac{4.8\pi
A\,\Gamma}{(m^2_{f_0}-m^2)^2+m^2_{f_0}\Gamma^2}+\Delta N_{BG}\,,
\end{equation}
where $\Delta N_{BG}=C_0+C_1m+C_2m^2$ represents a smooth
background and the parameter $A$ is the production of the
two-photon width $\Gamma_{f_0\to\gamma\gamma}$, branching ratio
$B(f_0\to\pi^+\pi^-)$, and known factors connected with the
detection efficiency and the setup luminosity \cite{MBC}. The
Belle Collaboration plans to report the information on
$\Gamma_{f_0\to\gamma\gamma}$ after the investigation of the
systematic error sources \cite{MBC}. The Belle data for the
$\gamma\gamma\to\pi^+\pi^-$ reaction cross section,
$\sigma(\gamma\gamma\to\pi^+\pi^-)$, in the region
$|\cos\theta^*|<0.6$, where $\theta^*$ is the center-of-mass
scattering angle of pion, with indication only statistical errors
are shown in Fig. 1(b). The comparison of these data with those of
the previous Mark II \cite{E1} and CELLO \cite{E2} experiments is
presented in Fig. 1(c).

It should be noted that, according the Belle data, the $f_0(980)$
resonance manifests itself in the $\gamma\gamma\to\pi^+\pi^-$
reaction cross section rather as a jump, or a step, with a width
of about 15 MeV and a height of about 11 nb, than as a clear peak;
see Fig. 1(b). In connection with this ``observation", as well as
bearing in mind some theoretical reasons (see below), we would
like to draw attention, especially of the experimentalists, to the
fact that Eq. (1) cannot be used to determine the physical
characteristics of the $f_0(980)$ resonance from the data on the
reaction $\gamma\gamma\to\pi^+\pi^-$. First, due to the proximity
of the $f_0(980)$ resonance to the $K\bar K$ thresholds and its
strong coupling to the $K\bar K$ channels, the propagator of the
form $1/(m^2_{f_0}-m^2-im_{f_0}\Gamma)$, with the total width
independent of $m$, cannot be applied in principle to the
description of the $f_0(980)$ resonance shape. Second, owing to
the $K^+K^-$ loop mechanism, the two-photon width of the
$f_0(980)$ resonance is a sharply varying function of $m$ just in
the $f_0(980)$ peak region. Therefore, it cannot be approximated
by a constant. And third, one cannot but take into account that
the $f_0(980)$ resonance strongly interferes with the considerable
$S$ wave background contributions in the
$\gamma\gamma\to\pi^+\pi^-$ reaction cross section.

In the present paper we analyze in detail the role of basic
dynamical mechanisms of the reaction $\gamma\gamma\to\pi^+\pi^-$
in the 1 GeV region and elucidate a possible form of the
$f_0(980)$ resonance manifestation in this channel. In so doing,
we tried to use sufficiently simple, but adequate to the highly
not simple physical situation, formulae free of unknown
parameters.

The paper is organized as follows. In Sec. II, the $K^+K^-$ loop
mechanism of the $f_0(980)\to\gamma\gamma$ decay is discussed.
This mechanism not only ensures the appreciable distortion of the
\begin{figure}
\includegraphics{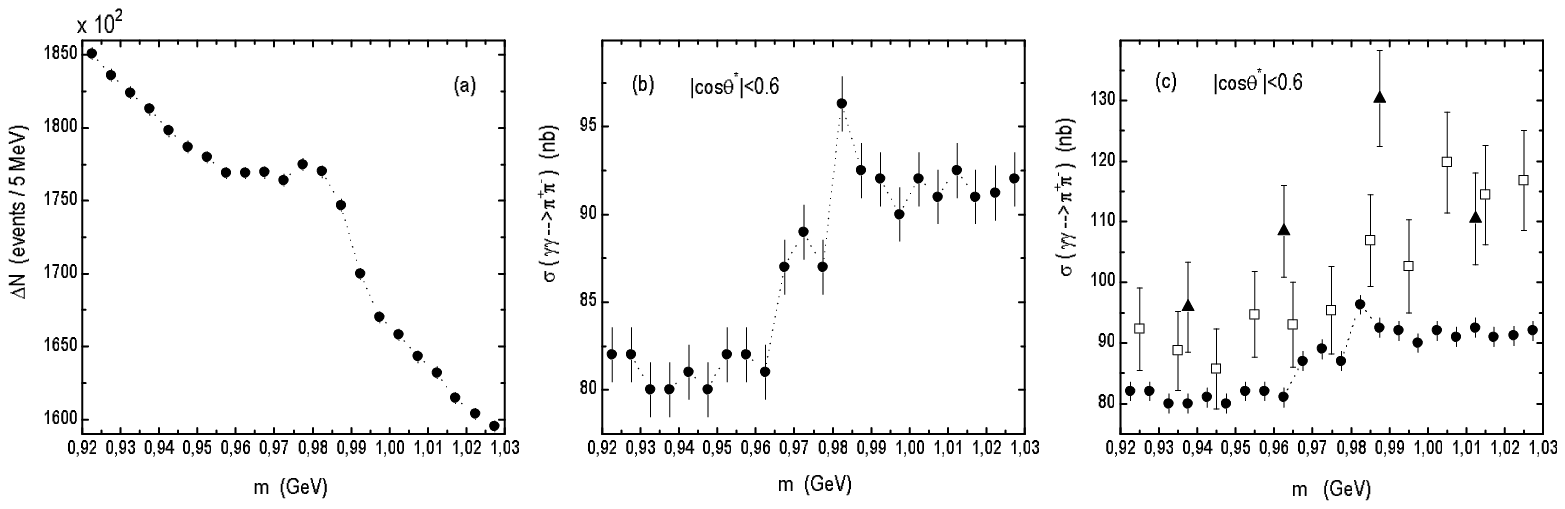}
\caption{(a) The Belle data \cite{MBC} for the invariant mass
distribution of $e^+e^-\to e^+e^-\pi^+\pi^-$ and $e^+e^-\to
e^+e^-\mu^+\mu^-$ events. (b) The Belle data \cite{MBC} for
$\sigma(\gamma\gamma\to\pi^+\pi^-)$; the quoted errors are
statistical only. (c) The comparison of the Belle data \cite{MBC}
for $\sigma(\gamma\gamma\to\pi^+\pi^-)$ with the analogous data of
the Mark II \cite{E1} (open squares) and CELLO \cite{E2} (full
triangles) experiments. The Belle points (full circles) are joined
by dotted lines for clearness.}
\end{figure}
$f_0(980)$ resonance shape in the reaction
$\gamma\gamma\to\pi^+\pi^-$ but also automatically yields a
reasonable estimate for the absolute magnitude of the $f_0(980)$
production cross section in this channel with the values of the
$f_0(980)$ resonance parameters compatible with the data on the
other reactions. Thus, we get good reasons to consider the
$K^+K^-$ loop mechanism as a major one of the $f_0(980)$
production in $\gamma\gamma$ collisions. In Sec. III, a simplest
dynamical model for the $S$ wave amplitude of the reaction
$\gamma\gamma\to\pi^+\pi^-$ in the 1 GeV region is examined and
the character of the interference between the background and
$f_0(980)$ resonance contributions, and thereby a possible
resulting shape of the $f_0(980)$ in the
$\gamma\gamma\to\pi^+\pi^-$ channel, is clarified. Most of all the
shape obtained resembles a step. This conclusion is supported by
the Belle data. A possible manifestation of the $f_0(980)$
resonance in the $\gamma\gamma\to\pi^0\pi^0$ channel is briefly
discussed. The general remarks and conclusions based on the
results of our analysis are formulated in Sec. IV.

\section{\boldmath{$K^+K^-$} loop mechanism of the
\lowercase{\boldmath{$f_0(980)\to\gamma\gamma$}} decay}

Perhaps, none of the known hadronic resonances can ``boast" of
such a variety of the forms of its own manifestation that the
$f_0(980)$ resonance possesses. The $f_0(980)$ shape in the
two-pion decay channel depends in a crucial way on the reaction
and varies from dips to peaks. In many respects this is due to the
fact that background contributions, usually accompanying the
$f_0(980)$ resonance, strongly change in passing from reaction to
reaction, which leads in its turn to the change of the
interference patterns in the resonance region. But, the even more
impressive thing is that there exist reactions in which the
$f_0(980)$ production amplitude itself sharply changes just in the
$f_0(980)$ peak region. First of all such a phenomenon takes place
in the radiative decays $\phi\to f_0(980)\gamma\to\pi\pi\gamma$
\cite{AI,AG1,AG2}. As predicted theoretically in Ref. \cite{AI}
and confirmed in the experiments performed at Novosibirsk
\cite{E7,E8} and Frascati \cite{E9}, these decays are determined
by the $K^+K^-$ loop mechanism of the $f_0(980)$ production,
$\phi\to K^+K^-\gamma\to f_0(980)\gamma\to\pi\pi\gamma$, the
amplitude of which is large, owing to the strong coupling of the
$f_0(980)$ to $K\bar K$, and changes very rapidly as a function of
two-pion invariant mass near the $K^+K^-$ threshold. The related
decay $\phi\to a_0^0(980)\gamma\to\eta\pi^0\gamma$ is also
determined by the $K^+K^-$ loop mechanism
\cite{AI,AG1,E8,E10,E11,AK1}. It should be also recalled that the
important role of this mechanism in the process $\gamma\gamma\to
a_0^0(980)\to\eta\pi^0$ was shown long ago in Ref. \cite{AS1}. The
above mentioned manifestations of the $K^+K^-$ loop mechanism
present important physical evidences in favor of the four-quark
($q^2\bar q^2$) nature of the $f_0(980)$ and $a_0^0(980)$
resonances \cite{AI,AS1,A1,A11}.

The presentation of high quality data from the Belle Collaboration
on the reaction $\gamma\gamma\to\pi^+\pi^-$ provides good reason
to discuss in detail the role of the $K^+K^-$ loop mechanism of
the $f_0(980)$ resonance production in $\gamma\gamma$ collisions.
As we shall show, it is very important, if not determining at all.
Note that the process $\gamma\gamma\to K\bar K\to
f_0(980)\to\pi\pi$ seems to be first mentioned in Ref. \cite{
ADS1}.

Thus, let us consider the shape of the $f_0(980)$ resonance
produced in the reaction $\gamma\gamma\to\pi^+\pi^-$ via the
$K^+K^-$ loop mechanism. This mechanism corresponds to the
following sequence of transitions. At first, there takes place the
formation of the $K^+K^-$ pair in $\gamma\gamma$ collisions, with
the amplitude which near the $K^+K^-$ threshold can be taken in
the Born approximation. Then, the $K^+K^-$ system turns into the
$f_0(980)$ resonance decaying further into $\pi^+\pi^-$. According
this prescription, the corresponding resonant contribution to the
$\gamma\gamma\to\pi^+\pi^-$ reaction cross section can be written
as
\begin{equation}\sigma_{f_0}(\gamma\gamma\to\pi^+\pi^-)
=\frac{8\pi}{m^2}\,\frac{m\Gamma^{Born}_{f_0\to
K^+K^-\to\gamma\gamma}(m)\,m\Gamma_{f_0\to
\pi^+\pi^-}(m)}{|D_{f_0}(m)|^2}\,.\end{equation} Here
\begin{equation} \Gamma^{Born}_{
f_0\to K^+K^-\to\gamma\gamma}(m)=\frac{1}{16\pi m}|M^{Born}_{
f_0\to K^+K^-\to\gamma\gamma}(m)|^2=\frac{\alpha^2}{4\pi^2}\,
|I_{K^+K^-}(m)|^2\,\frac{g^2_{f_0K^+K^-}}{16\pi m}
\end{equation} is the width of the $f_0(980)\to\gamma\gamma$ decay
due to the Born $K^+K^-$ loop mechanism, where
$\alpha=e^2/4\pi\approx1/137$ and the function $I_{K^+K^-}(m)$ is
\cite{AS1}
\begin{eqnarray} I_{K^+K^-}(m)=\left\{\begin{array}{ll}
\frac{m^2_{K^+}}{m^2}\left[\pi+i\ln\frac{1+\rho_{K^+}(m)}
{1-\rho_{K^+}(m)}\right]^2-1\,,\ \ & m\geq2m_{K^+}\,, \\
\frac{m^2_{K^+}}{m^2}[\pi-2\arctan|\rho_{K^+}(m)|]^2-1\,,\ \ &
0\leq m\leq 2m_{K^+}\,. \end{array}\right.
\end{eqnarray} The propagator of the $f_0(980)$ resonance with a mass
$m_{f_0}$ appearing in Eq. (2) has the form \cite{ADS2}
\begin{equation}\frac{1}{D_{f_0}(m)}=\frac{1}{m^2_{f_0}-m^2+
\sum_{{a\bar a}}[\mbox{Re}\Pi^{a\bar a}_{f_0}(m_{f_0})-\Pi^{a\bar
a}_{f_0}(m)]}\,,\end{equation} where $\Pi^{a\bar a}_{f_0}(m)$ is
the polarization operator of the $f_0(980)$ resonance
corresponding to the contribution of the $a\bar a$ intermediate
state ($a\bar a=\pi^+\pi^-,\ \pi^0\pi^0,\ K^+K^-,\ K^0\bar K^0$).
For $m\geq2m_a$,
\begin{equation}
\Pi^{a\bar a}_{f_0}(m)=\xi_{a\bar a}\frac{g^2_{f_0a\bar
a}}{16\pi}\rho_{a}(m)
\left[i-\frac{1}{\pi}\ln\frac{1+\rho_{a}(m)}{1-\rho_{a}(m)}\right]\,,
\end{equation}
$\rho_{a}(m)=(1-4m^2_a/m^2)^{1/2}$ [if $0\leq m\leq2m_a$, then
$\rho_{a}(m)\to i|\rho_{a}(m)|$], $\,\Gamma_{f_0\to a\bar
a}(m)=\mbox{Im}\Pi^{a\bar a}_{f_0}(m)/m= \xi_{a\bar a
}\,g^2_{f_0a\bar a}\,\rho_{a}(m)/16\pi m$ is the width of the
$f_0(980)\to a\bar a$ decay, here $\xi_{a\bar a}=1$, if $a\neq
\bar a$, and $\xi_{a\bar a}=1/2$, if $a=\bar a$, and $\
g^2_{f_0\pi^+\pi^-}=g^2_{f_0\pi^0\pi^0}=2g^2_{f_0\pi\pi}/3$, $\
g^2_{f_0K^+K^-}=g^2_{f_0K^0\bar K^0}=g^2_{f_0K\bar K}/2$, where $
g_{f_0\pi \pi}$ and $g_{f_0K\bar K}$ are the coupling constants of
the $f_0(980)$ to the $\pi\pi$ and $K\bar K $ channels,
respectively. Since we are interested in the $m$ region near the
$K\bar K$ thresholds, we take into account the $K^+$ and $K^0$
meson mass difference.

As for the $f_0(980)$ resonance parameters, the available data,
together with various model parametrizations, allow wide intervals
for their possible values; for example,
$m_{f_0}\approx(0.965-0.99)$ GeV,
$g^2_{f_0\pi\pi}/16\pi\approx(0.065-0.3)$ GeV$^2$, and
$g^2_{f_0K\bar K}/16\pi\approx(0.3-1.6)$ GeV$^2$, with the
preferred coupling-constant-squared ratio $R=g^2_{f_0K\bar
K}/g^2_{f_0\pi\pi}\approx4-6$, are quite compatible with the data
on most reactions of the $f_0(980)$ production
\cite{AI,AG1,E7,E8,E9,Fla,MOS,ADS2,AS2,CF,Abl,PDG}. For further
estimates and illustrations of the role of the $K^+K^-$ loop
mechanism, we use, in fact, all the range of possible values of
the $f_0(980)$ parameters.

Let us now discuss two most important features of the $K^+K^-$
loop mechanism which immediately follow from the above formulae.
First, as is seen from Fig. 2(a), the factor
$m\Gamma^{Born}_{f_0\to K^+K^-\to\gamma\gamma}(m)$ in Eq. (2)
sharply decreases just below the $K^+K^-$ threshold, i.e.,
directly in the $f_0(980)$ resonance region. For instance, it
falls relative to the maximum at $m=2m_{K^+}\approx0.9873$ GeV by
a factor of 1.69, 2.23, 2.75, 3.27, and 6.33 at $m=\,$0.98, 0.97,
0.96, 0.95, and 0.9 GeV, respectively. Such a behavior of
$m\Gamma^{Born}_{f_0\to K^+K^-\to\gamma \gamma}(m)$ strongly
suppresses the left wing of the $f_0(980)$ resonance peak defined
by $1/|D_{f_0}(m)|^2$ in Eq. (2). Second, from Eqs. (2) and (3) it
follows that for the $K^+K^-$ loop mechanism the magnitude of
$\sigma_{f_0}(\gamma\gamma\to\pi^+\pi^-)$ near the maximum,
located between $m_{f_0}$ and $2m_{K^+}$, is controlled mainly by
the parameter $R=g^2_{f_0K\bar K}/g^2_{f_0\pi\pi}$ and the value
of the function $|I_{K^+K^-}(m)|^2$. For example, if
$m_{f_0}<2m_{K^+}$, then, at $m=m_{f_0}$,
$\sigma_{f_0}(\gamma\gamma\to\pi^+\pi^-)=\alpha^2R|I_{K^+K^-}(m_{f_0})
|^2/[\pi m_{f_0}^2\rho_\pi(m_{f_0})]$. Furthermore, at fixed
$m_{f_0}$ and $R$, the $f_0(980)$ resonance shape in
$\sigma_{f_0}(\gamma\gamma\to\pi^+\pi^-)$ is very insensitive to
the absolute values of the coupling constants
$g^2_{f_0\pi\pi}/16\pi$ and $g^2_{f_0K\bar K}/16\pi$. As an
illustration we represent in Figs. 2(b) and 2(c) the cross section
$\sigma_{f_0}(\gamma\gamma\to\pi^+\pi^-)$ for four different sets
of the $f_0(980)$ resonance parameters: $m_{f_0}=0.98$ GeV, $R=4$,
$g^2_{f_0K\bar K}/16\pi=0.4$ GeV$^2$, and 1.2 GeV$^2$ (sets A and
B), and $m_{f_0}=0.97$ GeV, $R=5.33$, $g^2_{f_0K\bar
K}/16\pi=0.533$ GeV$^2$, and 1.6 GeV$^2$ (sets C and D). For sets
A and D the cross section smoothed with a Gaussian mass
distribution with the dispersion of 5 MeV (which we have chosen to
be equal to the $m$ step in the Belle experiment) is shown in
these figures for completeness. Multiplying the resulting cross
section values by a factor 0.6 [in accordance with the fact that
the data for $\sigma(\gamma\gamma\to\pi^+\pi^-)$ correspond to the
region $|\cos\theta^*|<0.6$], we obtain that, owing to the
$K^+K^-$ loop mechanism, the $f_0(980)$ resonance can manifest
itself in the measured $\gamma\gamma\to\pi^+\pi^-$ reaction cross
section at the level of about $15.5-17.5$ nb at the maximum. As is
clear from Fig. 1(b), this estimate for the scale of the
enhancement due to the $f_0(980)$ resonance contribution to the
$\pi^+\pi^-$ production cross section is in reasonable (if not
excellent) agreement with the Belle data. Thus, we conclude that
the $K^+K^-$ loop mechanism, which actually results from the
unitarity condition, can be primarily responsible for the
$f_0(980)$ resonance coupling to photons.

\begin{figure}
\includegraphics{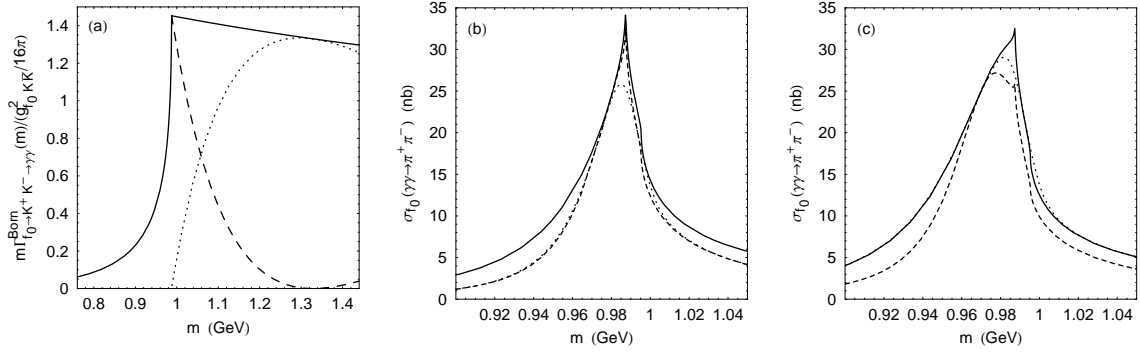}
\caption{(a) The solid curve shows $m\Gamma^{Born}_{ f_0\to
K^+K^-\to\gamma\gamma}(m)/(g^2_{f_0K\bar K}/16\pi)$ as a function
of $m$, see Eqs. (3) and (4); the dashed and dotted curves above
the $K^+K^-$ threshold correspond to the contributions of the real
and imaginary parts of the $M^{Born}_{f_0\to
K^+K^-\to\gamma\gamma}(m)$ amplitude to $m\Gamma^{Born}_{ f_0\to
K^+K^-\to\gamma\gamma}(m)$, respectively. The dashed, solid curves
in plots (b) and (c) show the cross section
$\sigma_{f_0}(\gamma\gamma\to\pi^+\pi^-)$ corresponding to the
$f_0(980)$ production via the $K^+K^-$ loop mechanism, see Eq.
(2), for the $f_0(980)$ parameter sets A, B and C, D,
respectively. For completeness, the dotted curves in (b) and (c)
show the examples of the cross sections smoothed with a Gaussian
mass distribution with the dispersion of 5 MeV for sets A and D,
respectively.}\end{figure}

It is clear that in there is no sense in speaking about a
two-photon width at the resonance point if the two-photon decay
width of the resonance varies rapidly within its hadronic width,
see Fig. 2(a). For the $K^+K^-$ loop mechanism, it is of interest
to evaluate the $f_0(980)\to\gamma\gamma$ width averaged by the
resonance mass distribution in the $\pi\pi$ channel,
$\langle\Gamma^{Born}_{f_0\to K^+K^-\to\gamma\gamma}
\rangle_{\pi\pi}$ \cite{AS1}. By definition,
\begin{equation}\langle\Gamma^{Born}_{f_0\to K^+K^-\to\gamma\gamma}
\rangle_{\pi\pi}=\int\limits_{m_1}^{m_2}\Gamma^{Born}_{f_0\to
K^+K^-\to\gamma\gamma}(m)\frac{3}{2}\left[\frac{m\Gamma_{f_0\to
\pi^+\pi^-}(m)}{\pi\,|D_{f_0}(m)|^2}\right]2mdm=\frac{3}{2}
\int\limits_{m_1}^{m_2}\frac{m^2}{4\pi^2}\sigma_{f_0}(\gamma\gamma\to
\pi^+\pi^-)dm\end{equation} [see also Eq.(2)]. This averaged width
can serve as an adequate, working characteristic of the $f_0(980)$
coupling to $\gamma\gamma$. Substituting
$\sigma_{f_0}(\gamma\gamma\to\pi^+\pi^-)$, shown in Figs. 2(b) and
2(c) in Eq. (7) and integrating, for example, over two $m$ regions
0.93 GeV $\leq m\leq$ 1.03 GeV and $2m_\pi\leq m<\infty$, we
obtain $\langle\Gamma^{Born}_{f_0\to
K^+K^-\to\gamma\gamma}\rangle_{\pi\pi}\approx0.114$ keV and 0.191
keV, respectively, for set A, 0.132 keV and 0.351 keV for set B,
0.129 keV and 0.211 keV for set C, and 0.152 keV and 0.377 keV for
set D. Defining also $\langle\Gamma^{Born}_{f_0\to
K^+K^-\to\gamma\gamma}\rangle_{K\bar K}$ in a similar way, we find
that the total averaged width of the $f_0\to\gamma\gamma$ decay
$\langle\Gamma^{Born}_{f_0\to K^+K^-\to\gamma\gamma}\rangle=
\langle\Gamma^{Born}_{f_0\to
K^+K^-\to\gamma\gamma}\rangle_{\pi\pi}+\langle\Gamma^{Born}_{f_0\to
K^+K^-\to\gamma\gamma}\rangle_{K\bar K}\approx0.14$ keV and 0.359
keV for the two above mentioned integration regions, respectively,
for set A, 0.164 keV and 0.884 keV for set B, 0.158 keV and 0.439
keV for set C, and 0.189 keV and 1.094 keV for set D. It is worth
to point out for comparison that $\Gamma^{Born}_{f_0\to
K^+K^-\to\gamma\gamma}(m)$ at the maximum, i.e.,
$\Gamma^{Born}_{f_0\to K^+K^-\to\gamma\gamma}(2m_{K^+})$, is
approximately equal to 0.589, 1.766, 0.785, and 2.355 keV for
$g^2_{f_0K\bar K}/16\pi$ from sets A, B, C, and D, respectively.

Certainly, the real situation in the $\gamma\gamma\to\pi^+\pi^-$
channel is more complicated because the $f_0(980)$ resonance in
this channel is by no means a solitary one. It is accompanied by
the considerable coherent background, and therefore the
interference effects are of great significance. Their role will be
analyzed in detail below in Sec. III.

We wish to conclude this section with a general remark concerning
the $f_0(980)$ resonance propagator, see Eq. (5), which we utilize
throughout here. As is shown in Ref. \cite{AK2}, this propagator
rigorously satisfies the K\"allen-Lehmann representation, i.e., it
possesses the analytic properties required in field theory. Thus,
the resonance mass distributions calculated with the use of this
propagator are automatically normalized to the corresponding
branching ratios of the $f_0(98 0)\to a\bar a$ decays, the sum of
which is exactly equal to unit, i.e.,
$$\int\limits_{2m_a}^{\infty}\left[\frac{m\Gamma_{f_0\to a\bar a
}(m)}{\pi\,|D_{f_0}(m)|^2}\right]2mdm=B(f_0(980)\to a\bar
a)\,,\qquad\sum_{a\bar a}B(f_0(980)\to a\bar a)=1\,.$$

\section{\boldmath{$S$} wave in the reaction
\boldmath{$\gamma\gamma\to\pi^+\pi^-$} near 1 GeV}

Let us consider a simplest dynamical model for the $S$ wave
amplitude of the reaction $\gamma\gamma\to\pi^+\pi^-$ in the 1 GeV
region. There are no arbitrary, free parameters in this model
(that is the parameters which would be unknown from other
reactions), and within its framework the character of the
interference between the background and $f_0(980)$ resonance
contributions, and thus a possible resulting $f_0(980)$ shape in
the $\gamma\gamma\to\pi^+\pi^-$ channel, will be fully elucidated.
The results obtained in this way will be useful, in particular, as
to estimate the potentialities and ``price"\ of the more
complicated model constructions.

Using the conventional normalization, we write the $S$ wave cross
section of the reaction $\gamma\gamma\to\pi^+\pi^-$, together with
the corresponding amplitude $A_S(m)$, in the form:
\begin{equation}\sigma_S(\gamma\gamma\to\pi^+\pi^-)= \frac{\rho_{\pi}
(m)}{32\pi m^2}|A_S(m)|^2\,,\end{equation}
\begin{eqnarray}A_S(m)
=M^{Born}_{\gamma\gamma\to\pi^+\pi^-}(m) +8\alpha
I_{\pi^+\pi^-}(m)\,T_{\pi^+\pi^-\to\pi^+\pi^-}(m) +8\alpha
I_{K^+K^-}(m)\,T_{K^+K^- \to\pi^+\pi^-}(m)\,.\end{eqnarray} Here
\begin{equation}M^{Born}_{\gamma\gamma\to\pi^+\pi^-}(m)=\frac{16\pi\alpha m^2
_\pi}{m^2\rho_\pi(m)}\,\ln\frac{1+\rho_\pi(m)}{1-\rho_\pi(m)}=\frac
{8\alpha}{\rho_\pi(m)}\mbox{Im}I_{\pi^+\pi^-}(m)
\end{equation} is the $S$ wave Born amplitude of the process
$\gamma\gamma\to\pi^+\pi^-$, the function $I_{\pi^+\pi^-}(m)$
results from Eq. (4) by replacing $m_{K^+}$ and $\rho_{K^+}(m)$ by
$m_{\pi}$ and $\rho_\pi(m)$, respectively, and
$T_{\pi^+\pi^-\to\pi^+\pi^-}(m)$ and $T_{K^+K^- \to\pi^+\pi^-}(m)$
are the $S$ wave amplitudes of hadronic reactions indicated in
their subscripts. Hence it is obvious that the second and third
terms on the right-hand side of Eq. (9) correspond to the
contributions from the $\gamma\gamma\to\pi^+\pi^-$ and
$\gamma\gamma\to K^+K^-$ Born amplitudes modified by the final
state interactions. Such a structure of the amplitude $A_S(m)$ can
be easily obtained within the framework of the field-theoretical
model in which the electromagnetic Born amplitudes are the only
primary sources of the $\pi^+\pi^-$ and $K^+K^-$ pairs, and the
strong amplitudes, used for unitarization of the Born
contributions, are constructed by summing up all the $s$ channel
bubble diagrams. In so doing, the strong amplitudes can involve,
in principle, any number of resonances plus background
contributions to describe the relevant data on the phase shifts
and inelasticities. The resulting strong and electromagnetic
amplitudes in such a model are unitary. This model has a very old
history \cite{ZGMZ,Tir} and up to now was successfully used,
together with its dispersive modifications, as the effective tool
in analyzing dynamics of electromagnetic and strong interaction
processes, see for example \cite{AG2,BG,Men,Joh,GRR,AS3,AS4}.

The amplitude $T_{\pi^+\pi^-\to\pi^+\pi^-}(m)$ is related to the
phase shifts $\delta^I_0(m)$ and inelasticities $\eta^I_0(m)$ of
the $S$ wave $\pi\pi$ scattering amplitudes with definite isospin
$I=0,2$ in the conventional way: $T_{
\pi^+\pi^-\to\pi^+\pi^-}(m)=\frac{2}{3}T^0_0(m)+\frac{1}{3}
T^2_0(m)$, where
$T^I_0(m)=\{\eta^I_0(m)\exp[2i\delta^I_0(m)]-1\}/[2i\rho_\pi(m)]$.
As is well known, the only, strongly coupled $S$ wave channels in
the 1 GeV region are the $\pi\pi$ and $K\bar K$ channels with
$I=0$. Therefore we set $\eta^2_0(m)=1$ for all $m$ of interest
and $\eta^0_0(m)=1$ for $m<2m_{K^+}$. Then, for $m<2m_{K^+}$, the
amplitude $A_S(m)$ can be rewritten as, see Eqs. (9) and (10),
\begin{eqnarray} A_S(m)=e^{i\delta^0_0(m)}\{A_{S,0}(m)+
A_{S,2}(m)\cos[\delta^2_0(m)-\delta^0_0(m)]+iA_{S,2}(m)\sin[\delta^2_0(m)
-\delta^0_0(m)]\}\,,
\end{eqnarray} where the amplitudes $A_{S,I}(m)$ with $I=0$ and
2 have the form:
\begin{eqnarray}& A_{S,0}(m)
=\frac{2}{3}M^{Born}_{\gamma\gamma\to\pi^+\pi^-}
(m)\cos\delta^0_0(m) & \nonumber \\ & +[8\alpha/\rho_\pi(m)]\,
\mbox{Re}[I_{\pi^+\pi^-}(m)]\frac{2}{3}\sin\delta^0_0(m)+8\alpha
I_{K^+K^-}(m)T_{K^+K^- \to\pi^+\pi^-}(m)\,e^{-i\delta^0_0(m)}\,, &
\end{eqnarray}
\begin{eqnarray}& A_{S,2}(m)=\frac{1}{3}\{M^{Born}_{\gamma\gamma\to\pi^+\pi^-}
(m)\cos\delta^2_0(m)+[8\alpha/\rho_\pi(m)]\,
\mbox{Re}[I_{\pi^+\pi^-}(m)]\sin\delta^2_0(m)\}\,.&
\end{eqnarray} Because for $m<2m_{K^+}$ the
imaginary part of the function $I_{K^+K^-}(m)$ vanishes, see Eq.
(4), and the phase of the amplitude $T_{K^+K^- \to\pi^+\pi^-}(m)$
reduces to $\delta^0_0(m)+n\pi$ (where $n=0$ or 1) in accordance
with unitarity, it is easy to see that all the terms in the
amplitudes $A_{S,0}(m)$ and $A_{S,2}(m)$ are real.
\begin{figure}
\includegraphics{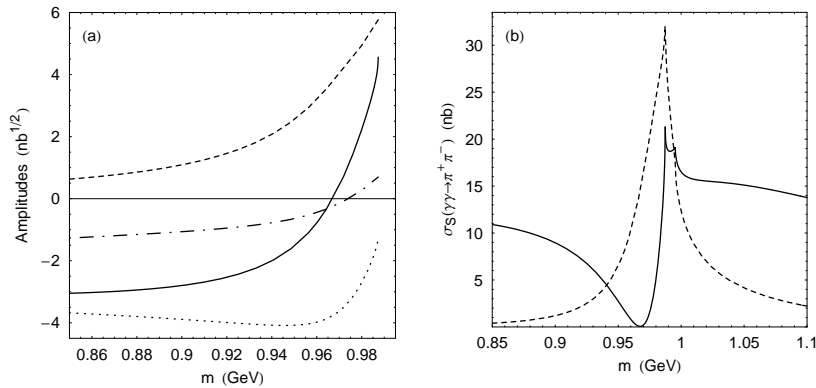}
\caption{(a) The components of the amplitude
$\sqrt{\rho_{\pi}(m)/(32\pi m^2)}\,e^{-i\delta^0_0(m)}A_S(m)$ for
$m\leq2m_{K^+}$ are shown; see Eqs. (8), (9), (11)\,--\,(13). The
solid curve corresponds to the real part of this amplitude, which
is added up of the $f_0(980)$ production amplitude due to the
$K^+K^-$ loop mechanism, shown by the dashed curve, and the real
part of the sum of the background amplitudes, sown by the dotted
curve. The dot-dashed curve corresponds to the imaginary part of
this amplitude, which is stipulated only by the $I=2$
contribution. (b) The solid curve shows the cross section
$\sigma_S(\gamma\gamma\to\pi^+\pi^-)$ calculated by using Eqs. (8)
and (9). The dashed curve shows the contribution to this cross
section caused by the $f_0(980)$ production only via the $K^+K^-$
loop mechanism. A comparison of these curves gives, in particular,
a good idea of the important role of the interference between the
background and resonance contributions. The values of the
parameters utilized in constructing the curves in (a) and (b)
correspond to set A, see the text.}\end{figure} Moreover, all of
these terms have well definite signs. Begin with the amplitude
$T_{K^+K^- \to\pi^+\pi^-}(m)=T_{\pi^+\pi^-\to K^+K^-}(m)$ in Eq.
(12). Its sign, $(-1)^n$, is known experimentally and it is
positive \cite{MOS,Mor,Wet,Est,Coh}. In terms of the $f_0(980)$
coupling constants this means that if we parametrize the amplitude
$T_{K^+K^- \to\pi^+\pi^-}(m)$ in the 1 GeV region as
\cite{MOS,ADS2,AS2}
\begin{equation}T_{K^+K^-\to\pi^+\pi^-}(m)=
\frac{g_{f_0\pi^+\pi^-} g_{f_0K^+K^-}}{16\pi
D_{f_0}(m)}\,e^{i\delta_B(m)}\,,\end{equation} where $\delta_B(m)$
is a smooth and large phase (of about 90$^\circ$ for $m\approx1$
GeV) of the elastic background in the $I=0$ $S$ wave $\pi\pi$
channel, then the production $g_{f_0\pi^+\pi^-}g_{f_0K^+K^-}$ is
positive \cite{MOS,Mor,Est}. Recall that with such a
parametrization the $\pi\pi$ scattering amplitude $T^0_0(m)$ has
the form \cite{Fla,MOS,ADS2,AS2}:
\begin{equation}T^0_0(m)=\frac{\eta^0_0(m)e^{2i\delta^0_0(m)}-1}
{2i\rho_\pi(m)}=\frac{1}{\rho_\pi(m)}\left[\frac{e^{2i\delta_B(m)}
-1}{2i}+e^{2i\delta_B(m)}\,\frac{m\Gamma_{f_0\pi\pi}(m)}
{D_{f_0}(m)}\right]\,, \end{equation} and that the $f_0(980)$
resonance appears as a dip in $|T^0_0(m)|$ \cite{fn1}. Equations
(14) and (15) will be used in the following. Thus, the last term
on the right-hand side of Eq. (12) is positive because, according
Eq. (4), $I_{K^+K^-}(m)>0$ for $0<m\leq2m_{K^+}$. Now we take into
account the following circumstances. For
$0.85$\,GeV$\,<m<2m_{K^+}$, the phase shift $\delta^0_0(m)$
increases with $m$ from 90$^\circ$ to about 200$^\circ$ sharply
flying up near the $K^+K^-$ threshold; see, for example, Ref.
\cite{Hya}. In the same region of $m$, the phase shift
$\delta^2_0(m)$ is of about $-\mbox{(19--24)}^\circ$; see, for
example, Ref. \cite{Hoo}. Moreover,
$\mbox{Re}[I_{\pi^+\pi^-}(m)]<0$ for $m>0.376$ GeV. So, for
$0.85$\,GeV$\,<m<2m_{K^+}$, the first term on the right-hand side
of Eq. (12) is negative, the second term is also negative at least
up to 0.98 GeV, and it is small in magnitude for
$0.98$\,GeV$\,<m<2m_{K^+}$. Finally, it is easy to check that the
amplitude $A_{S,2}(m)\cos[\delta^2_0(m)-\delta^0_0(m)]$, see Eqs.
(11) and (13), is also negative for $0.85$\,GeV$\,<m<2m_{K^+}$.

Thus, one can conclude that, for $m<2m_{K^+}$, the sharply
increasing with $m$, $f_0(980)$ production amplitude due to the
$K^+K^-$ loop mechanism has to interfere destructively with the
accompanying background contributions in
$\sigma_S(\gamma\gamma\to\pi^+\pi^-)$. Such an interference is
able to suppress the left wing of the $f_0(980)$ resonance
practically in full. A detailed illustration of the described
general picture is presented in Fig. 3. In constructing the curves
shown in this figure, we used set A for the values of the
$f_0(980)$ resonance parameters and approximated the smooth phase
shifts $\delta_B(m)$ and $\delta^2_0(m)$ by the following
expressions:
$\delta_B(m)=\rho_\pi(m)\sum_{n=0}^3q^{2n}_\pi(m)a_{2n}=
\rho_\pi(m)[0.1243+q^2_\pi(m)16.32-
q^4_\pi(m)73.50+q^6_\pi(m)118.3]$ and
$\delta^2_0(m)=q_\pi(m)b_0/[1+\sum_{n=1}^3q^{2n}_\pi(m)b_{2n}]=
q_\pi(m)0.9098/[1+q^2_\pi(m)2.629-
q^4_\pi(m)13.19+q^6_\pi(m)18.83]$, where $\delta_B(m)$ and
$\delta^2_0(m)$ in radians and $q_\pi(m)=m\rho_\pi(m)/2$ in units
of GeV. Note that in this way we obtain the excellent description
of the $S$ wave $\pi\pi$ scattering data \cite{Hya,Hoo,Ros} at
least in the $m$ region from $2m_\pi$ up to 1.2 GeV (for example,
according to our fit, the $S$ wave $\pi\pi$ scattering length
$a^0_0\approx0.229/m_\pi$). In Fig. 3(a) the solid curve shows
that the real part of the amplitude $\sqrt{\rho_{\pi}(m)/(32\pi
m^2)}\,e^{-i\delta^0_0(m)}A_S(m)$, see Eqs. (8), (9), and
(11)\,--\,(13), vanishes at $m\approx0.967$ GeV as a result of the
compensation of the resonance and background contributions. As is
seen from Fig. 3(b), this leads to a minimum in the cross section
at the place of the left wing of the $f_0(980)$ resonance. As a
whole, the resulting cross section
$\sigma_S(\gamma\gamma\to\pi^+\pi^-)$ near 1 GeV in Fig. 3(b)
resembles a step. Furthermore, we verified that sets B, C, and D
for the $f_0(980)$ resonance parameters yield very similar results
for $A_S(m)$ and $\sigma_S(\gamma\gamma\to\pi^+\pi^-)$.

To compare the model with the data pertaining to the partial solid
angle, one must yet take into account the interference of the
amplitude $A_S(m)$ with the higher partial waves. Usually, the
measurements of the reaction $\gamma\gamma\to\pi^+\pi^-$ are
performed in the angular region $|\cos\theta^*|<Z_0<1$. The
$\gamma\gamma\to\pi^+\pi^-$ cross section is presented as the sum
of the cross sections
$\sigma_{\lambda=0}(\gamma\gamma\to\pi^+\pi^-,|\cos\theta^*|<Z_0)$
and
$\sigma_{|\lambda|=2}(\gamma\gamma\to\pi^+\pi^-,|\cos\theta^*|<Z_0)$,
where $\lambda$ is a photon helicity difference. In the $Z_0<1$
case, all the partial waves interfere between themselves in both
cross sections. The cross section with $|\lambda|=2$ is dominated
by the $D$ wave Born contribution and the well known $f_2(1270)$
resonance \cite{MBC,E1,E2,Joh}. The $f_2(1270)$ coupling to the
$\gamma\gamma$ system in the $\lambda=0$ state is small
\cite{E1,Joh}. Therefore, we assume for estimate that in the 1 GeV
region all the higher partial waves with $\lambda=0$ are defined
simply by the corresponding $\gamma\gamma\to\pi^+\pi^-$ Born
amplitude. Then, $
\sigma_{\lambda=0}(\gamma\gamma\to\pi^+\pi^-,|\cos\theta^*|<Z_0)$
can be written in the form:
\begin{eqnarray} & \sigma_{\lambda=0}(\gamma\gamma\to\pi^+\pi^-,
|\cos\theta^*|<Z_0)=\frac{\rho_{\pi} (m)}{32\pi
m^2}\Biggl\{Z_0|\widetilde{A}_S(m)|^2+C\,\mbox{Re}[\widetilde{A}_S(m)]
\Biggr.& \nonumber \\ & \Biggl.\times
\frac{1}{\rho_\pi(m)}\,\ln\frac{1+Z_0\rho_\pi(m)}{1-Z_0\rho_\pi(m)}+
C^2\left[\frac{Z_0/2}{1-Z^2_0\rho^2_\pi(m)}+\frac{1}{4\rho_\pi(m)}
\,\ln\frac{1+Z_0\rho_\pi(m)}{1-Z_0\rho_\pi(m)}\right]\Biggr\}\,, &
\end{eqnarray} where the amplitude
$\widetilde{A}_S(m)=A_S(m)-M^{Born}_{
\gamma\gamma\to\pi^+\pi^-}(m)$, see Eq. (9), and $C=32\pi\alpha
m^2_\pi/m^2$. With the use of Eq. (16), one can easily verify that
for the typical value of $Z_0=0.6$ the higher partial wave
influence, certainly, exists, but it is not too large.
\begin{figure}
\includegraphics{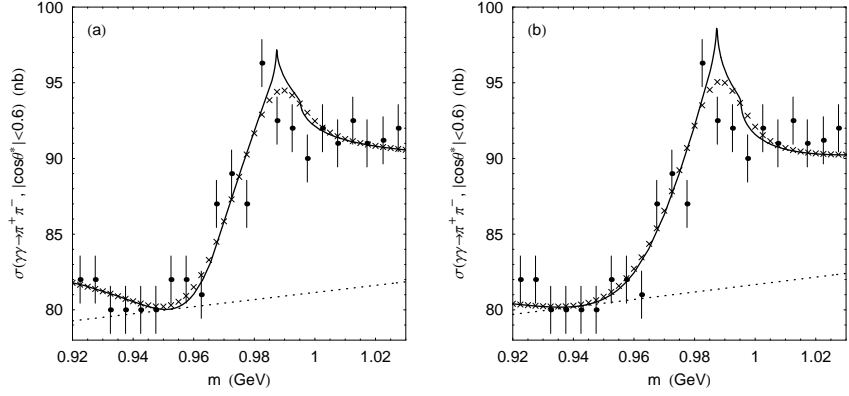}
\caption{The comparison of the model predictions for the cross
section $\sigma(\gamma\gamma\to\pi^+\pi^-,|\cos\theta^*|<0.6)=
\sigma_{\lambda=0}(\gamma\gamma\to\pi^+\pi^-,|\cos\theta^*|
<0.6)+\sigma_{|\lambda|=2}(\gamma\gamma\to\pi^+\pi^-,|\cos\theta^*|
<0.6)$ with the Belle data [from Fig. 1(b)] in the $f_0(980)$
resonance region. (a) The solid curve, crosses, and dotted line
show the cross section without and with a Gaussian mass smearing
(with the dispersion of 5 MeV), and the contribution of the the
$|\lambda|=2$ cross section approximated by a linear function of
$m$, respectively. The presented fit [obtained, in particular,
with use of Eqs. (9), (10), and (14)\,--\,(16)] corresponds to the
model parameters from set E, see the text. (b) The same as in (a)
but for the model parameters from set F.}\end{figure}

Figures 4(a) and 4(b) illustrate the comparison of the model
predictions for
$\sigma(\gamma\gamma\to\pi^+\pi^-,|\cos\theta^*|<0.6)=
\sigma_{\lambda=0}(\gamma\gamma\to\pi^+\pi^-,|\cos\theta^*|
<0.6)+\sigma_{|\lambda|=2}(\gamma\gamma\to\pi^+\pi^-,|\cos\theta^*|
<0.6)$ with the Belle data in the $f_0(980)$ resonance region. To
obtain the curves in Fig. 4(a), we performed the simultaneous fit
to the Belle data \cite{MBC} and the well known $S$ wave $\pi\pi$
scattering data from Refs. \cite{Hya,Ros}. In so doing, we used
Eqs. (9), (10), and (14)\,--\,(16), the above mentioned expression
for $T^2_0(m)$, and the approximation of the cross section with
$|\lambda|=2$ by a linear function of $m$, $C_1+C_2m$ [of course,
this is a reasonable approximation only in the considered, narrow
region of $m$ around the $f_0(980)$ resonance]. The parameters
obtained (set E) are $m_{f_0}=0.9676$ GeV,
$g^2_{f_0\pi\pi}/16\pi=0.07017$ GeV$^2$, $g^2_{f_0K\bar
K}/16\pi=0.3442$ GeV$^2$ (R=4.9), $C_1=57.69$ nb, $C_2=23.45$
nb/GeV, and $a_{2n=0,2,4,6}=0.1404$, 17.17, $-80.17$, 127.4,
respectively. In order to illustrate that the Belle data tolerate,
in fact, the wide range for the $f_0(980)$ coupling constant
values, we fixed $g^2_{f_0K\bar K}/16\pi=1.6$ GeV$^2$ and
performed once again the fit to the above mentioned data. For this
case, the parameters obtained (set F) are $m_{f_0}=0.968$ GeV,
$g^2_{f_0\pi\pi}/16\pi=0.2438$ GeV$^2$ (R=6.56), $C_1=57.05$ nb,
$C_2=24.62$ nb/GeV, and $a_{2n=0,2,4,6}=0.01903$, 18.13, $-96.71$,
173.2, respectively, and the resulting picture is shown in Fig.
4(b). As a whole, we obtain the quite satisfactory, qualitative
agreement with the data in both the magnitude and shape of the
$f_0(980)$ resonance manifestation. The strong difference of the
$f_0(980)$ resonance shape in the $\gamma\gamma\to\pi^+\pi^-$
reaction cross section from the shape of the solitary Breit-Wigner
resonance is a result of fine interference effects between the
different contributions. As we have made sure, the considered
dynamical model provides a fairly good basis for understanding
these effects. The model unambiguously points to the destructive
interference pattern between the resonance and background
contributions in the $m$ region below the $K^+K^-$ threshold.

Now we discuss, in brief, a possible manifestation of the
$f_0(980)$ resonance in the $S$ wave $\gamma\gamma\to\pi^0\pi^0$
reaction cross section. In the considered model we have:
\begin{equation}\sigma_S(\gamma\gamma\to\pi^0\pi^0)= \frac{\rho_{\pi}
(m)}{64\pi m^2}|B_S(m)|^2\,,\end{equation}
\begin{equation} B_S(m)=8\alpha
I_{\pi^+\pi^-}(m)\,T_{\pi^+\pi^-\to\pi^0\pi^0}(m)+8\alpha
I_{K^+K^-}(m)\,T_{K^+K^-\to\pi^0\pi^0}(m)\,,\end{equation}
\begin{figure}
\includegraphics{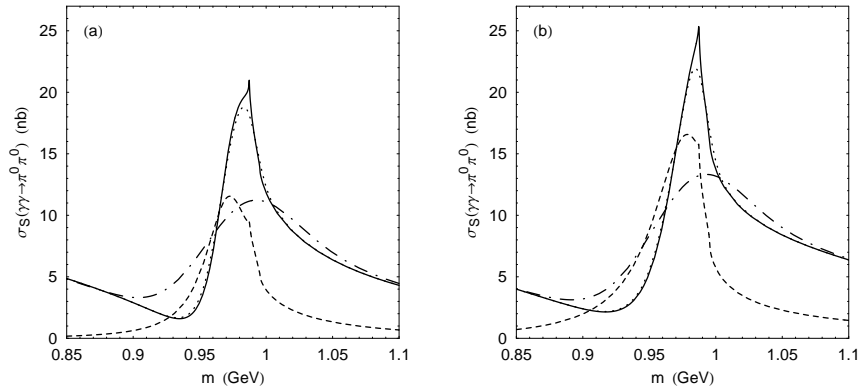}
\caption{(a) The solid curve shows the cross section
$\sigma_S(\gamma\gamma\to\pi^0\pi^0)$ calculated with use of Eqs.
(17) and (18) in the case of the model parameters from set E. The
dotted and dot-dashed curves show the same cross section but
smoothed with a Gaussian mass distribution with the dispersion of
5 and 30 MeV, respectively. The dashed curve shows the
contribution caused by the $f_0(980)$ resonance production via the
$K^+K^-$ loop mechanism only. (b) The same as in (a) but for the
model parameters from set F.}\end{figure} where
$T_{\pi^+\pi^-\to\pi^0\pi^0}(m)=\frac{2}{3}T^0_0(m)-\frac{2}{3}
T^2_0(m)$ and
$T_{K^+K^-\to\pi^0\pi^0}(m)=T_{K^+K^-\to\pi^+\pi^-}(m)$. In
comparison with the amplitude $A_S(m)$, see Eq. (9), the amplitude
$B_S(m)$ does not contain the Born term and the $T^2_0(m)$
amplitude contribution is doubled and has the opposite sign. These
differences are  essential. As is seen from Fig. 5, the $f_0(980)$
resonance in the $\gamma\gamma\to\pi^0\pi^0$ channel has to
manifest itself as a distinct peak. In this respect, the reaction
$\gamma\gamma\to\pi^0\pi^0$, generally speaking, is more preferred
than the reaction $\gamma\gamma\to\pi^+\pi^-$. Unfortunately, in
the Crystal Ball \cite{E4,E5} and JADE \cite{E6} experiments, the
$\gamma\gamma\to\pi^0\pi^0$ cross section was scanned with a
50-MeV and 30-MeV-wide step, respectively. Such a mass resolution
is still lacking to discover the $f_0(980)$ peak. Figures 5(a) and
5(b) show, in particular, that a Gaussian smearing with the
dispersion of 30 MeV leaves nothing from the specific features of
the $f_0(980)$ peak in $\sigma_S(\gamma\gamma\to\pi^0\pi^0)$.
Notice, that there are no contradictions between the presented
estimate for the smoothed $\sigma_S(\gamma\gamma\to\pi^0\pi^0)$
and the normalized Crystal Ball data \cite{E4,E5} for
$\sigma(\gamma\gamma\to\pi^0\pi^0,|\cos\theta^*|<0.8,0.7)$.

Of course, the considered model allows us to predict the $S$ wave
$\gamma\gamma\to\pi\pi$ reaction cross sections for a more wide
region of $m$ than the neighborhood of the $f_0(980)$ resonance.
The corresponding cross sections
$\sigma_S(\gamma\gamma\to\pi^+\pi^-)$ and
$\sigma_S(\gamma\gamma\to\pi^0\pi^0,|\cos\theta^*|<0.8)$ in the
$m$ region from $2m_\pi$ to 1.2 GeV are shown in Fig. 6 for the
model parameters corresponding to sets E and F. Unfortunately, in
such a wide $m$ interval we cannot directly compare the
predictions for $\sigma_S(\gamma\gamma\to\pi\pi)$ with experiment,
because this requires the accurate $S$ wave data obtained by
separating highest partial waves with the use of a partial wave
analysis of the reaction events. For example, in the reaction
$\gamma\gamma\to\pi^+\pi^-$, the $D$ wave contribution with
$|\lambda|=2$ can constitute from 75\% to 90\% of the total cross
section for $m>0.5$ GeV. In the $\gamma\gamma\to\pi^0\pi^0$
cannel, the $D$ wave contribution, caused in the main by the
$f_2(1270)$ resonance, is also very important for $m>0.85$ GeV, as
is clear from Fig. 6(c). Hence, the thorough separation of the
large $D$ wave background is of crucial importance for the
extraction of the $S$ wave in both reactions.

Finally, we wish to say a few words about ambiguities which, in
fact, inevitably occur in theoretical models for the amplitudes of
electromagnetic interactions of hadrons. Concretely, we bear in
mind rather evident possibilities of the incorporation of some
unknown, free parameters into the aforesaid model. One of these
parameters is the so-called direct $f_0(980)\to\gamma\gamma$
coupling constant, $g^0_{f_0\to\gamma\gamma}$. Taking account of
this constant, the corresponding total amplitude of the
$f_0\to\gamma\gamma$ decay can be written as
$M_{f_0\to\gamma\gamma}(m)=M^{Born}_{f_0\to K^+K^-
\to\gamma\gamma}(m)+g^0_{f_0\to\gamma\gamma}$, where
$M^{Born}_{f_0\to K^+K^- \to\gamma\gamma}(m)$ is the amplitude due
to the Born $K^+K^-$ loop mechanism from Eq. (3). Of course, for
any mechanism, the two-photon decay amplitude of any scalar meson
must be proportional to $m^2$ for $m\to0$, as, for example, the
Born amplitude $M^{Born}_{f_0\to K^+K^-\to\gamma\gamma}(m)$.
However, if we are interested in only the narrow $m$ region around
1 GeV, the adding of the constant $g^0_{f_0\to\gamma\gamma}$ to
$M^{Born}_{f_0\to K^+K^-\to\gamma\gamma}(m)$ is a quite reasonable
approximation. About the coupling constant
$g^0_{f_0\to\gamma\gamma}$ one can say as follows. It can have
neither the value comparable in magnitude and coincident in sign
with the value of $M^{Born}_{f_0\to K^+K^-\to\gamma\gamma}(m)$ at
the maximum, i.e., with $M^{Born}_{f_0\to K^+K^-
\to\gamma\gamma}(2m_{K^+})=\alpha(\pi^2/4-1)g_{f_0K^+K^-}/2\pi$,
nor the value comparable in magnitude but opposite in sign with
$M^{Born}_{f_0\to K^+K^-\to\gamma\gamma}(2m_{K^+})$, since
otherwise the $\gamma\gamma\to\pi^+\pi^-$ reaction cross section
in the $f_0(980)$ region would be in sharp contradiction with the
data, in both magnitude and shape. Moreover, there are no evidence
for the presence of the pointlike $f_0(980)\phi\gamma$ interaction
from the data on the $\phi\to\pi\pi\gamma$ decays
\cite{E7,E8,E9,A11}. Actually, experiment tells us that the direct
$f_0\to\gamma\gamma$ coupling seems to be small. Any reliable
theoretical estimates for this coupling have not existed yet.
Serious experimental and theoretical search for its signs together
with those of the direct coupling of the $f_0(600)/\sigma$ to
$\gamma\gamma$ are still a matter of the future.

\begin{figure}
\includegraphics{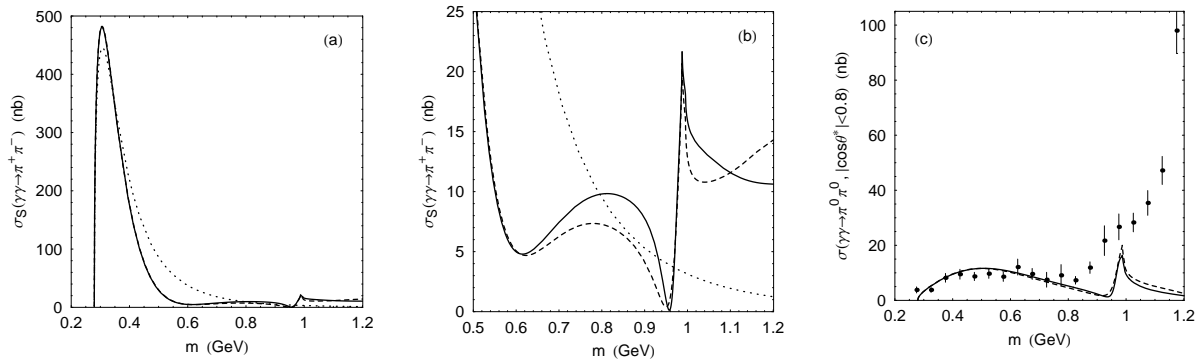}
\caption{The $S$ wave cross sections
$\sigma_S(\gamma\gamma\to\pi^+\pi^-)$, in (a) and (b), and
$\sigma_S(\gamma\gamma\to\pi^0\pi^0,|\cos\theta^*|<0.8)$, in (c),
obtained in the considered model for the wide regions of $m$. They
are shown by the solid and dashed curves corresponding to the
model parameters from sets E and F, respectively. The dotted
curves in (a) and (b) correspond to the $S$ wave Born
contribution. The experimental points in (c) are the Crystal Ball
data on $\sigma(\gamma\gamma\to\pi^0\pi^0,|\cos\theta^*|<0.8)$
\cite{E4} (the quoted errors are statistical only); the rise of
the measured cross section for $m>0.8$ GeV is due to the
$f_2(1270)$ resonance contribution \cite{E4}.}\end{figure}

\section{Conclusion}

The present analysis was stimulated by the Belle data \cite{MBC}.
The main results consist in the following.

(i) It has been shown that the $K^+K^-$ loop mechanism provides
the absolutely natural and reasonable scale of the $f_0(980)$
resonance manifestation in the $\gamma\gamma\to\pi^+\pi^-$ and
$\gamma\gamma\to\pi^0\pi^0$ reaction cross sections.

(ii) It has been shown that the shape of the $f_0(980)$ resonance
in the reaction $\gamma\gamma\to\pi^+\pi^-$ has nothing to do with
the shape of a solitary Breit-Wigner resonance. This result is
supported by the Belle data. In so doing, the observed pattern of
the $f_0(980)$ peak distortion can be easily explained with use of
the simple dynamical model.

Certainly, for the more full understanding of the situation, the
information based on a partial wave analysis of the
$\gamma\gamma\to\pi^+\pi^-$ reaction events in the $f_0(980)$
resonance region would be extremely useful. The huge statistics
collected in the Belle experiment \cite{MBC}, in principle, allows
one to hope for the successful performance of such an analysis.

It is clear from the preceding discussion that high quality data
on the reaction $\gamma\gamma\to\pi^0\pi^0$ would be also highly
desirable, because the relative role of the background
contributions in the $f_0(980)$ region in this channel is
considerably smaller than in the charged one.

The new stage of high statistics measurements of the processes
$\gamma\gamma\to\pi^+\pi^-$,\, $\gamma\gamma\to\pi^0\pi^0$,\,
$\gamma\gamma\to\eta\pi^0$,\, $\gamma\gamma\to K^+K^-$, and
$\gamma\gamma\to K^0\bar K^0$, begun by the Belle Collaboration,
undoubtedly, will serve the further progress of physics of light
scalar mesons.

\begin{center}{\bf ACKNOWLEDGMENTS}\end{center}

This work was supported in part by the Presidential Grant No.
2339.2003.2 for the support of Leading Scientific Schools.


\begin{thebibliography}{99}
\bibitem{fn1} Here it should be noted that the $\pi^+\pi^-$ intermediate
state contribution to the $f_0(980)\to\gamma\gamma$ decay width
makes no sense by itself in contrast to the $K^+K^-$ one. This is
so because the contributions to the $\gamma\gamma\to\pi^+\pi^-$
amplitude $A_S(m)$ from the $f_0(980)$ resonance production  via
the $\pi^+\pi^-$ intermediate state and the smooth, large elastic
$\pi^+\pi^-$ background cancel each other completely at $m\approx
m_{f_0}$ in the second term in Eq. (9), see also Eqs. (11), (12),
and (15) [$\delta^0_0(m)\approx180^\circ$ at this point].
Nevertheless, as described below, the the wings of this term
outside of its minimal value, together with the Born term in Eq.
(9), strongly interfere in the cross section with the third
resonance term from Eq. (9).
\end{thebibliography}
\end{document}